\documentclass[aps,pra,reprint,groupedaddress]{revtex4-1}

\usepackage{graphicx}% Include figure files

\begin{document}

%\preprint{}

\title{Curvature Capillary Repulsion}


%\author{Iris B. Liu}
%\thanks{}
%\affiliation{Chemical and Biomolecular Engineering Department, University of Pennsylvania}
\author{Iris B. Liu}
\altaffiliation
{Department of Chemical and Biomolecular Engineering, University of Pennsylvania, Philadelphia}

\author{Giulia Bigazzi}
\altaffiliation
{Department of Chemical Engineering, Universit\`{a} di Pisa, Pisa, Italy}

\author{Nima Sharifi-Mood}
\altaffiliation
{Department of Chemical and Biomolecular Engineering, University of Pennsylvania, Philadelphia}

\author{Lu Yao}
\altaffiliation{Department of Chemical and Biomolecular Engineering, University of Pennsylvania, Philadelphia}

\author{Kathleen J. Stebe}
\altaffiliation
{Department of Chemical and Biomolecular Engineering, University of Pennsylvania, Philadelphia}
\email{kstebe@seas.upenn.edu}


\date{\today}

\begin{abstract}
Directed assembly of colloids is an exciting field in materials science to form structures with new symmetries and responses. Fluid interfaces have been widely exploited to make densely packed ordered structures. We have been studying how interface curvature can be used in new ways to guide structure formation. On a fluid interface, the area of the deformation field around adsorbed microparticles depends on interface curvature; particles move to minimize the excess area of the distortions that they make in the interface. For particles that are sufficiently small, this area decreases as particles move along principle axes to sites of high deviatoric curvature. We have studied this migration for microparticles on a curved host interface with zero mean curvature created by pinning an oil-water interface around a micropost. Here, on a similar interface, we demonstrate capillary curvature repulsion, that is, we identify conditions in which  microparticles migrate away from high curvature sites. Using theory and experiment, we discuss the origin of these interactions and their relationship to the particle's undulated contact line. We discuss the implications of this new type of interaction in various contexts from materials science to microrobotics. 
\end{abstract}

\pacs{Microassembly, particle trapping, }
%\keywords{}

\maketitle

\section{Introduction}

\begin{figure}
\centering
  \includegraphics[scale=1.1]{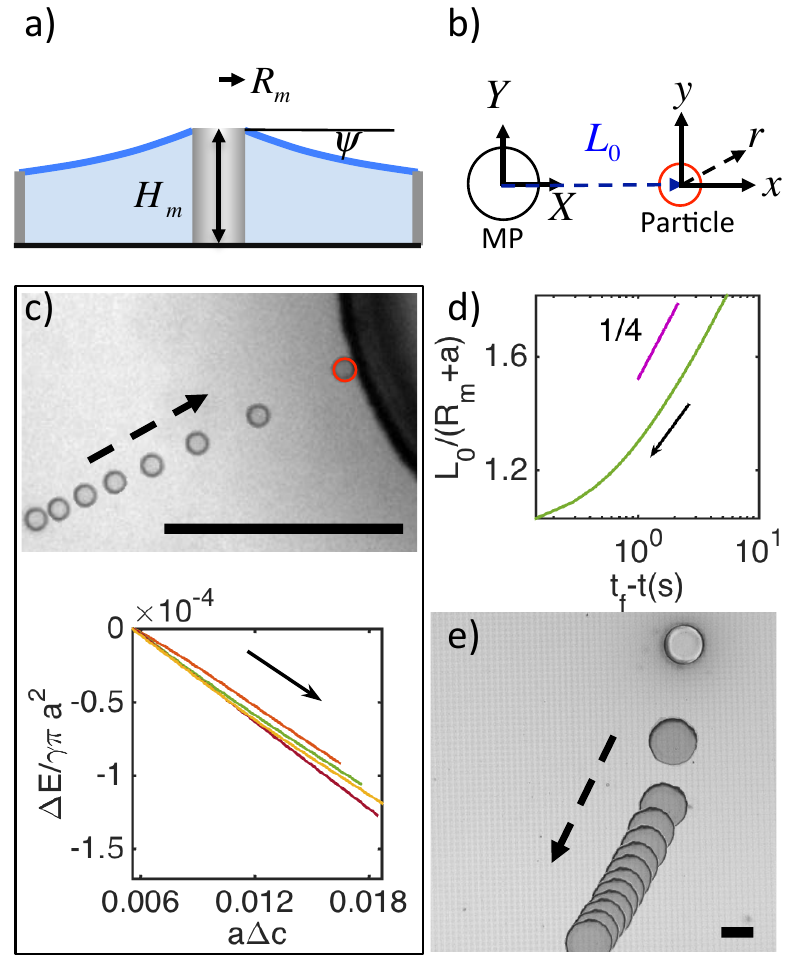}
  \caption[Curvature attraction and curvature repulsion]{\textbf{Curvature attraction and curvature repulsion}. Arrows indicate direction of migration. (a) Schematic of the interface shape formed by pinning fluid interface to the edge of a micropost and an outer ring.  (b) Coordinates at center of micropost and at the particle located a distance $L_0$ from the post center. (c) (Top) A microdisk (2$a$=10 $\mu$m) migrates to the edge of a micropost (250 $\mu$m diameter). (Bottom) The energy dissipated as such particles migrate to the high curvature regions is linear in deviatoric curvature. (d) The power law dependence confirms the dominance of the quadrupolar mode. (e) A disk (2$a$=150 $\mu$m) migrates away from the micropost (125 $\mu$m diameter).}
   \label{rep_fig_edge_eqm_rep}
\end{figure}

Fluid interfaces are excellent sites to organize microscale particles with potential to form new functional structures. Microparticles are firmly attached to fluid interfaces by significant trapping energies associated with the elimination of a small patch of interface by the particle adsorption. On planar interfaces, crystalline packing of microparticles can be achieved, e.g., by exploiting electrostatic repulsion and weak confinement in a gravitational well \cite{Pieranski}. More complex arrangements can be achieved by a number of routes. For example, if complex fluids like nematic liquid crystals are used as subphases, repulsion between the colloids owing to their topological defects can also form crystalline packings \cite{weishao}, and elastic energies in the nematic subphase can guide complex open structures to form \cite{CavallaroLCPNAS}. Particle shape can also dictate the symmetries of structures at fluid interfaces; the capillary energy, given by the product of surface tension and area, near a particle depends on features like particle aspect ratio and the presence of sharp edges. Thus, capillary interactions depend on particle shape and orientation, so particles assemble in preferred configurations, e.g., uncharged ellipsoids in side-to-side configurations \cite{Loudet, Vermant} and cylindrical microparticles in end-to-end configurations \cite{Lewand2008Langmuir, Botto2012Review, Botto2012SM}. In this communication, we discuss recent studies of capillary assembly at curved fluid interfaces,  introduce the new concept of curvature capillary repulsion, and conclude with a broader vision of how these interactions and analogous interactions in other soft matter hosts can be exploited in a number of contexts. \\
\indent On a curved fluid interface, the excess area around a colloid depends on the underlying interface shape. As a result, curvature can act as an ``external field"  to dictate particle behavior. In prior work, we have reported capillary curvature attraction in the limit of small particle radius to interface radius of curvature \cite{Cavallaro27122011, Yaodisk, Nimasphere, CMRstebe}. Here we report more complex behavior that emerges when this limit is relaxed. To situate this result, we first briefly recapitulate the main concepts that explain curvature capillary attraction. We then motivate our discussion of curvature capillary repulsion.\\
\indent The phenomenon of capillary curvature attraction is now well established, having been reported for microcylinders \cite{Cavallaro27122011}, microdisks \cite{Yaodisk}, and microspheres \cite{Nimasphere}. These studies were performed for particles on a host interface created by pinning a water interface around a micropost of height $H_m$, radius $R_m$, with interface slope at the micropost of $-\tan \Psi$, where $\Psi$ is roughly 15$^\circ$. A layer of hexadecane is gently placed above this water layer to minimize fluxes from evaporation. A schematic of this arrangement is shown in Fig. \ref{rep_fig_edge_eqm_rep}a. Particles are introduced to this oil super-phase sediment under gravity and attach to the interface, avoiding the use of spreading solvent. The interface slope is small, and all experiments are performed at distances from the center of the micropost $L$ small compared to the capillary length. Similar migrations are observed for cylinders, spheres and disks \cite{CMRstebe}. A typical trajectory for a disk migrating on the curved fluid interface is shown in Fig. \ref{rep_fig_edge_eqm_rep}c. The particle attaches to the interface at some distance from the micropost. It then migrates along paths radial to the micropost until contact. For this situation, the interface shape is governed by the Laplace equation. The host interface shape $h(L)$ in the absence of the particle is simply
\begin{eqnarray}
h(L)=H_m-R_m\tan\Psi \ln(\frac{L}{R_m})
\label{host}
\end{eqnarray}
The interface shape has zero mean curvature, and proves a versatile platform for investigation of curvature capillary effects. This expression can be expanded in a polar coordinate system $(r, \phi)$ defined in the plane tangent to the interface in powers of $\lambda=\frac{a}{L_0}$, where $a$ is the particle radius and $L_0$ is the distance from the post center where a particle adsorbs. To leading order, the local host interface height above this reference plane has the form:
\begin{eqnarray}
h^{0}(r,\phi)=\frac{\Delta c(L_0)}{4}r^2\cos2\phi
\end{eqnarray}
This is a saddle surface, characterized by the deviatoric curvature $\Delta c=\frac{1}{R_1}-\frac{1}{R_2}$, where $R_1$ and $R_2$ are the principal radii of curvature at $L_0$. When colloidal particles attach to the interface, they make distortions in the interface. Colloidal particles typically have pinned contact lines \cite{Kaz:2012pin, Manoharan:2015pin, ColosquiPRL2013}; these contours can be described in terms of an expansion in Fourier modes. Each mode excites an interfacial distortion given by the corresponding mode in a multipole expansion. For particles with negligible body forces and body torques, the leading order Fourier mode has quadrupolar symmetries \cite{Stamou, CMRstebe}, so the contact line, to leading order is described:
\begin{eqnarray}
h^{particle}(r=a, \phi)=h_2 \cos2(\phi+\alpha_2)
\end{eqnarray}
where $h_2$ is the magnitude of the quadrupolar mode, and the phase angle $\alpha_2$  is zero when the quadrupolar rise axis aligns with rise axis of the saddle surface.  The interfacial distortion associated with this contour couples with the interface curvature in a manner akin to a charged multipole in an external electric field. For a small particle \cite{Yaodisk, CMRstebe}, the capillary energy for a microparticle on such a curved interface differs from the case of a planar interface:
\begin{eqnarray}
E({{L}_{0}})={{E}_{planar}}-\gamma \pi {{a}^{2}}\frac{{{h}_{2}}\Delta c({{L}_{0}})}{2}\cos 2{{\alpha }_{2}}
\label{energy}
\end{eqnarray}
This expression predicts a torque and a force; it indicates that particles decrease their capillary energy by rotating the particle to align its quadrupolar rise along the rise axis of the host interface $({{\alpha }_{2}}=0)$, as has been observed for cylindrical microparticles \cite{Lewandowski2010,Cavallaro27122011}, and by migrating to sites of high curvature.   

This predicted linear dependence of the capillary energy on the deviatoric curvature at the particle center of mass can be compared to experiment like that presented in Fig. \ref{rep_fig_edge_eqm_rep} c-d. The Reynolds number Re= $\frac{av\rho}{\mu}\sim 10^{-5}$ for a typical trajectory, where $a$ is the characteristic length of the particle, $v$ is the particle velocity, $\rho$ is the fluid density, and $\mu$ is the fluid viscosity.  Thus, the particles migrate in creeping flow, and the energy dissipated along a trajectory can be extracted by integrating along a particle path. 
\begin{eqnarray}
\Delta E &&=\gamma \pi {{a}^{2}}\frac{{{h}_{2}}}{2}(\Delta c({{L}_{0}}_{,f})-\Delta c({{L}_{0}}_{,i})) \nonumber \\
&& =6\pi \mu a\int\limits_{{{s}_{i}}}^{{{s}_{f}}}{{{C}_{D}}vds}
\end{eqnarray}
We have performed this integration two ways.  We have truncated the particle trajectory $\sim10$ particle radii from the post to avoid hydrodynamic interactions with the wall, adopting Lamb's drag coefficient on a disk \cite{Lamb}. We have also integrated over the energy dissipated over the entire trajectory, adopting drag coefficients for disks near bounding surfaces in Fig. \ref{rep_fig_edge_eqm_rep}c  (details in Supplementary Material) \cite{trahan}. These graphs confirm the predicted linear dependence of the migration energy on the local deviatoric curvature, and allow the magnitude of ${{h}_{2}} \sim 71-101$ nm to be inferred. From (\ref{energy}), the capillary force can be found: 
\begin{eqnarray}
F_{cap}=\gamma\pi a^2 \frac{h_2}{2}\frac{d\Delta c}{dL_0}\delta_L
\end{eqnarray} 
A balance of this force and viscous drag predicts a power law dependence in $L_0$ vs. $(t_f-t)$, for particles far enough from the post to neglect hydrodynamic interactions, where $t_f$ is the time that the particle ceases its migration. This power law is apparent in the data (Fig. \ref{rep_fig_edge_eqm_rep}d). Note that the trajectories are non-Brownian; typical energy dissipated along a trajectory is of order $10^5 k_BT$. Typical capillary forces magnitudes are $\sim$ 10pN for a disk of radius 10 $\mu m$ for the range of $h_2$ inferred. The gravitational force on the particle can also be estimated \cite{Chan, nicolson_1949}. The particle migrates to minimize its potential energy, constrained to move along the interface: 
\begin{eqnarray}
{{E}_{g}}=\left[ ({{\rho }_{p}}-{{\rho }_{1}}){{\Phi }_{1}}+({{\rho }_{p}}-{{\rho }_{2}})(1-{{\Phi }_{1}}) \right]{{V}_{p}}gh
\end{eqnarray}
where ${{\rho }_{p}}$ is the particle density, ${{\rho }_{i}}$ is the density of water (i=1) and oil (i=2), and ${{\Phi }_{1}}$ is the volume fraction of the particle immersed in the aqueous phase, and $V_p$ is the volume of the particle. The corresponding force owing to gravity is:
\begin{eqnarray}
F_g=-[(\rho_P-\rho_1)\Phi_1+(\rho_P-\rho_2)(1-\Phi_1)]V_P g \frac {\delta h}{\delta z}
\end{eqnarray}
The ratio of the gravitational force to the capillary force is between $10^{-5}$ to $10^{-4}$. The negligible contributions by gravity is also reflected in the small value for the Bond number, a dimensionless number defines the contribution of gravitational forces to surface tension forces, $Bo= \frac{\Delta \rho gL^2}{\gamma}=1.3\times 10^{-6}$. \\
\indent What if we consider larger particles or smaller posts? What new phenomena emerge? Such a trajectory is shown in Fig. \ref{rep_fig_edge_eqm_rep}e for a disk 125 $\mu$m in diameter on an interface similar to that in Fig. \ref{rep_fig_edge_eqm_rep}c.  This larger particle moves along a roughly radial path \textit{away} from the post.  Here, we delve into the origin of this apparent curvature repulsion, and find a rich variety of behaviors. 
\section{Theory}
We have considered several potential sources of this repulsion. Gravitational forces on interfacially-trapped particles remain small, even for these larger particles.  Interactions with the micropost can also be considered.  Equation (\ref{energy}) was derived using the method of reflections \cite{brenner_1962, KimKarrila}, in which the micropost determines the host interface shape near the particle, and the disturbance made by the particle is determined assuming a pinned contact line. The curvature-dependent part of the energy associated with this disturbance is reported in equation (\ref{energy}). A similar calculation can be done to understand interactions of the particle-sourced disturbance with the micropost. However, the energy associated with these interactions is negligible, with largest contribution being two orders of magnitude smaller than the leading order term (see Supplemental Material). Finally, we consider the role of higher order Fourier modes in the particle's distorted contact line, which can couple with higher order modes in the local description of the host interface. Theory presented below shows that the importance of these modes increase with $\lambda$; we present a simple theory based on these interfacial details, and compare it favorably to the experiments.\\
\indent  Here, we derive a more general  expression for a particle interacting with the host interface shape. The entire host interface profile is described by (\ref{host}). When a microdisk attaches to this interface at some distance $L_0$ from the center of the post, the disturbance created by the particle decays monotonically over distances comparable to the particle radius $a$. To understand the energy associated with this deformation, we first find an expression for the local shape of the host interface at $L_0$.  In the limit of $\lambda$=$\frac{a}{L_0}<1$ we expand (\ref{host}) in a power series in a local coordinate $x,y,z$ with origin at $L_0$ (Fig. \ref{rep_fig_edge_eqm_rep}b), where the height is defined as the distance above the plane tangent to the interface. We perform a coordinate transformation, using the following relationship between the two coordinates, 
\begin{eqnarray}
  & {{L}^{2}}={{X}^{2}}+{{Y}^{2}} \\ 
 & X={{L}_{0}}+x \\ 
 & Y=y \\
 & Z=Z_0
\end{eqnarray}
where X, Y, Z are the coordinate centered at the bottom center of the micropost, $Z_0$ is the interface height at the particle center of mass, and $h^0$ is the height of the interface above $Z_0$. Expanding the natural log term in (\ref{host}), and scaling heights with $a$, the dimensionless interface profile is
\begin{eqnarray}
{{\widetilde{h}}^0}=&& \frac{1}{2}\frac{\lambda{{R}_{m}}\tan \psi }{{{L}_{0}}}({{\widetilde{x}}^{2}}-{{\widetilde{y}}^{2}})-\frac{{{\lambda}^{2}}{{R}_{m}}\tan \psi }{{{L}_{0}}}(\frac{{{\widetilde{x}}^{3}}}{3}-\widetilde{x}{{\widetilde{y}}^{2}}) \nonumber \\
&& +\frac{1}{4}\frac{{{\lambda}^{3}}{{R}_{m}}\tan \psi }{{{L}_{0}}}({{\widetilde{y}}^{4}}+{{\widetilde{x}}^{4}}-6{{\widetilde{x}}^{2}}{{\widetilde{y}}^{2}})+... 
\label{hostnd}
\end{eqnarray}
where ${\widetilde{h}^0}$=$\frac{h}{a}$, ${\widetilde{x}}$=$\frac{x}{a}$, ${\widetilde{y}}=\frac{y}{a}$. Equation (\ref{hostnd}) can be recast in a  polar coordinate $\widetilde{x}=\widetilde{r}\cos \phi $, $\widetilde{y}=\widetilde{r}\sin \phi $: 
\begin{eqnarray}
{{\widetilde{h}}^0}=&&\frac{1}{2}\frac{\lambda{{R}_{m}}\tan \psi }{{{L}_{0}}}{{\widetilde{r}}^{2}}\cos 2\phi -\frac{1}{3}\frac{{{\lambda}^{2}}{{R}_{m}}\tan \psi }{{{L}_{0}}}{{\widetilde{r}}^{3}}\cos 3\phi \nonumber \\
&&+\frac{1}{4}\frac{{{\lambda}^{3}}{{R}_{m}}\tan \psi }{{{L}_{0}}}{\widetilde{r}}^{4}\cos 4\phi +... 
\end{eqnarray}
Noting that  $\Delta c$=$\frac{2R_m tan \psi}{{L_0}^2}$ the local description of the host interface shape becomes,
\begin{eqnarray}
{h^0}=\frac{1}{4}\Delta c{{r}^{2}}\cos 2\phi -\frac{1}{6}\frac{\Delta c}{{{L}_{0}}}{{r}^{3}}\cos 3\phi +\frac{1}{8}\frac{\Delta c}{{{L}_{0}}^{2}}{{r}^{4}}\cos 4\phi  \nonumber +... \\
\end{eqnarray}
A circular disk on the fluid interface disturbs the interface height owing to its pinned contact line; the particle-sourced disturbance owing to the pinned, undulated contact line can be expressed in terms of a multipole expansion:
\begin{eqnarray}
{{h}^{particle}}=&&{{h}_{2}}\frac{{{a}^{2}}}{{{r}^{2}}}\cos 2(\phi+\alpha_2) +{{h}_{3}}\frac{{{a}^{3}}}{{{r}^{3}}}\cos 3(\phi+\alpha_3) \nonumber \\
&&+{{h}_{4}}\frac{{{a}^{4}}}{{{r}^{4}}}\cos 4(\phi+\alpha_4)+...
\end{eqnarray}
where the subscripts 2, 3, and 4 indicate mode of deformation, e.g., the quadrupolar, hexapolar, and octopolar modes of deformation and $\alpha_n$ is the phase angle for mode $n$. To find the interface shape for the particle attached to the interface, we note that the solution takes the form of a multipole expansion, and apply the pinning boundary condition, 
\begin{eqnarray}
h(r=a,\phi )={{h}^{particle}(r=a, \phi)} 
\end{eqnarray}
and require that, far from the particle, the interface recovers the host interface shape,
\begin{eqnarray}
\lim_{r \to \infty}h(r,\phi )={{h}^{0}}
\end{eqnarray}
The full expression of the resulting interface profile near the particle is
\begin{eqnarray}
h=&&h^0+{{h}^{particle}}-\frac{1}{4}\frac{\Delta c{{a}^{2}}}{{{r}^{2}}}{{a}^{2}}\cos 2(\phi+\alpha_2) \\ \nonumber
&& +\frac{1}{6}\frac{\Delta c}{{{L}_{0}}}\frac{{{a}^{3}}}{{{r}^{3}}}{{a}^{3}}\cos 3(\phi+\alpha_3) \\ \nonumber
&& -\frac{1}{8}\frac{\Delta c}{{{L}_{0}}^{2}}\frac{{{a}^{4}}}{{{r}^{4}}}{{a}^{4}}\cos 4(\phi+\alpha_4) +...
\end{eqnarray} 
The capillary energy associated with the particle on this interface is the product of the constant interfacial tension and the change in surface area owing to particle attachment.  The details of this calculation are reported in our prior work on the quadrupolar disturbance, so details are not given here. The capillary energy around the particle is: 
\begin{eqnarray}
\Delta E=&&\gamma \pi {{a}^{2}}\frac{\Delta {{c}}}{2}\left\{ -{{h}_{2}}\cos 2{{\alpha }_{2}}+\lambda {{h}_{3}}\cos 3{{\alpha }_{3}}\right.\nonumber \\
 &&\left. -{{\lambda }^{2}}{{h}_{4}}\cos 4{{\alpha }_{4}} +... \right\} + \text{self terms}
\label{manymodenergy}
\end{eqnarray}
where the self terms indicate interactions quadratic in the magnitude of each mode that would occur for particles on a planar interfaces, which are independent of interface shape.
\section{Experimental}
We perform two series of experiments. We study disks of diameter 25 $\mu m$ and 150 $\mu m$ and vary the micropost diameter while fixing its height and the slope of the interface at the micropost's edge. For each experiment, we compute the Bond number $Bo$; for these systems $Bo$ ranges from 1 $\times 10^{-5}$ to 6 $\times 10^{-5}$.
 \subsection{Particle fabrication}
Epoxy resin particles are fabricated using standard lithographic techniques. A negative tone photoresist, SU-8 (MicroChem Corp.), is spin-coated onto a chrome-sputtered silicon wafer. After soft baking at 95$^\circ$C, the photoresist is exposed to UV light on a tabletop mask aligner (OAI Model 100) through a mask with an array of circular holes.  The photoresist is cross-linked on a hot plate at 95$^\circ$C.  The sample is then developed in SU-8 developer solution to dissolve the unexposed region, leaving solid circular disks on the wafer. The disks are released from wafer by sonicating in chrome etchant. Subsequently, particles are cleaned, washed with water and isopropanol, and stored in hexadecane for further uses. The disk particles aspect ratio is 0.2, thickness to diameter.
\subsection{Molding the Interface}
We fabricate the vessel used to mold the fluid interface using lithographic techniques given in detail in a previous publication \cite{Cavallaro27122011}. In short, three layers of lithography are built on a silicon wafer: a wetting layer of an array of microcylinders of height 5 $\mu$m, a bounding ring of radius 1.27 cm and height 25 $\mu$m, and a micropost of height 250 $\mu$m. The micropost height $H_m$ in all the experiment is 200 $\mu$m. Microposts of varying radii are fabricated, including 47 $\mu$m, 57 $\mu$m, 125 $\mu$m, 250 $\mu$m, and 600 $\mu$m.
\subsection{Experimental Protocol}
We add water to the vessel to pin at the top edge of the micropost and at the edge of the bounding ring. The interface slope is determined by the volume of water; the volume is sufficiently large and the ring is sufficiently far from the micropost that the mean curvature of the interface is negligible. In these experiments, the pinning angle, $\psi$, is kept between 15$^\circ \leq \psi \leq$ 20$^\circ$. This is the maximum slope of the interface. We then carefully place a layer of hexadecane onto the water subphase to form a curved oil-water interface. We introduce particles to the interface by gently adding drops of microparticles dispersed in hexadecane to the superphase where they slowly sediment and attach to the interface. We study trajectories after this attachment event. The vessel is placed under an upright microscope (Zeiss M1m) in a reflective mode. Particle trajectories are recorded using a high resolution camera at rate of 0.141 or 0.200 ms per frame. 
\section{Results and Discussions}
\begin{figure}
\centering
  \includegraphics[scale=1]{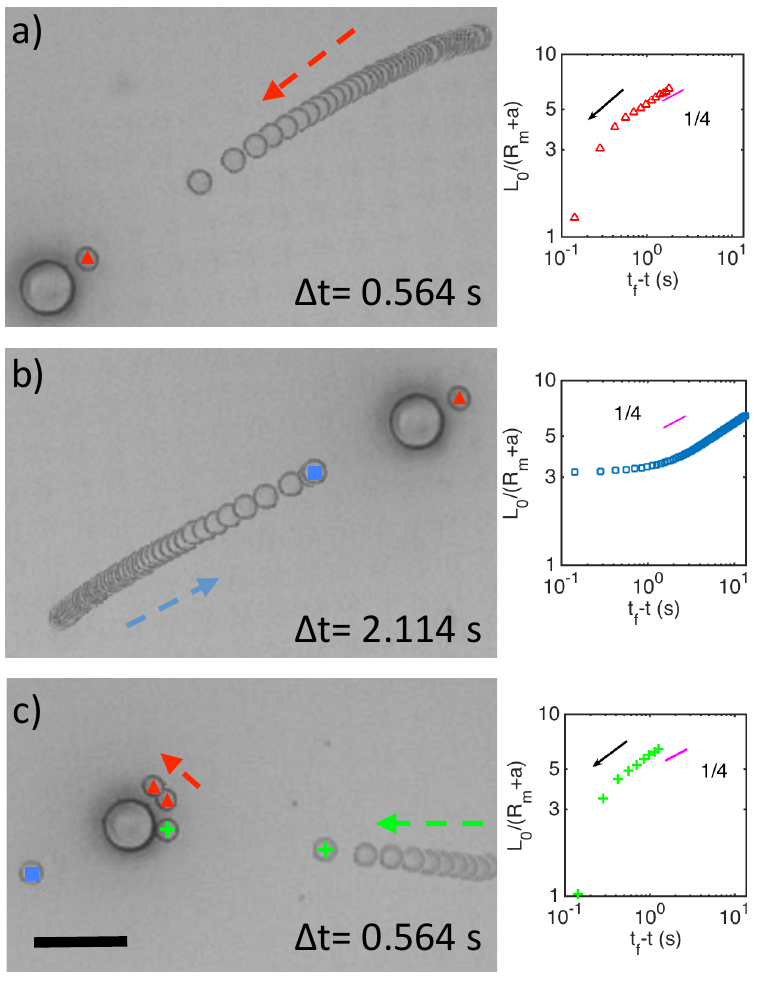}
  \caption[Bocce Ball: Trajectories of microdisks ($2a$=25 $\mu$m) around a circular micropost.]{\textbf{Bocce Ball: Trajectories of microdisks ($2a$=25 $\mu$m) around a circular micropost.} Left panel: Time stamped image of microdisks trajectory on a curved oil-water interface where the micropost is 57 $\mu$m in diameter. Right panel: log-log graph of $L_0/(R_m+a)$ vs $t_f-t$ for each trajectory. (a) A microdisk (red triangle) migrates along curvature gradients toward the micropost and stops before contacting the post (at 50.9 $\mu$m center-to-center distance). (b) A microdisk (blue square) migrate towards the micropost and stops at 128.0 $\mu$m, several particle radii from the post. (c) A microdisk (green cross) migrates towards the edge of the micropost.  This incoming disk interacts strongly with the first disk, causing it to move at a constant distance from the post to a new location, as such, its final position is shifted slightly. Scale bar is 100 $\mu$m.}
  \label{rep_fig_baci}
\end{figure}

\begin{figure}
\centering
  \includegraphics[scale=1]{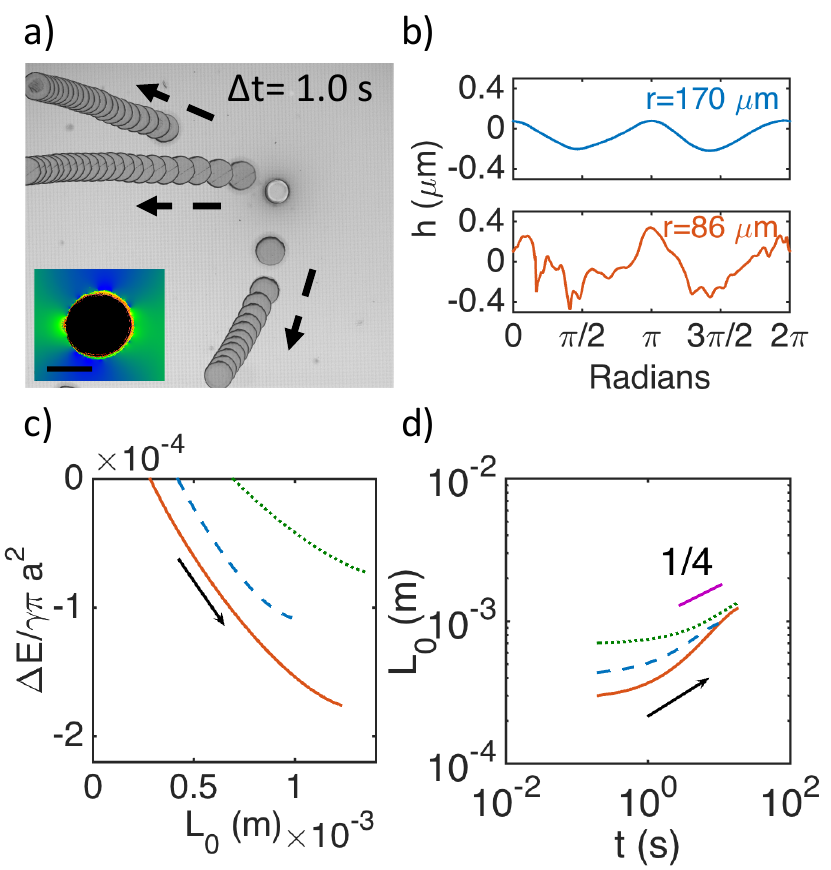}
  \caption[Curvature repulsion of microdisks ($2a$=150 $\mu$m) around circular microposts]{\textbf{Curvature repulsion of microdisks ($2a$=150 $\mu$m) around circular microposts}. Time stamped image of microdisks trajectories on curved oil-water interfaces. (a) Three microdisks are repelled by the curved oil-water interface and migrate radially outwards from the 125 $\mu$m diameter micropost. Inset: interferogram showing quadrupolar distortion far from the particle and puckered interface near the particle. Scale bar is 100 $\mu$m. (b) Top: quadrupolar undulation of interface height along a circle of radius 170 $\mu$ m from particle center. Bottom: rugged interface height along a circle of radius 86 $\mu$ m from particle center. (c) Energy dissipated along a trajectory. (d) $L_0$ vs. $t$: evidence of repulsive quadrupole in the far field.}
  \label{rep_fig_largep}
\end{figure}

\begin{figure}
\centering
  \includegraphics[scale=1.2]{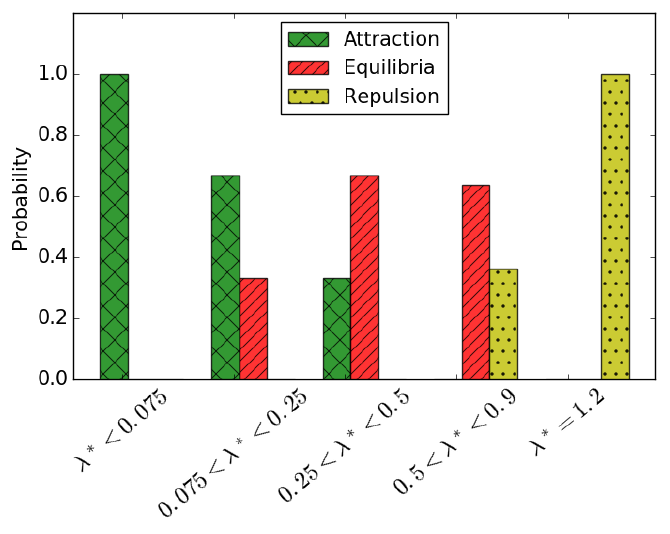}
  \caption[Transition from curvature attraction to repulsion with $\lambda^*$.]{\textbf{Transition from curvature attraction to repulsion with $\lambda^*$.} A histogram summarizing all of the particle trajectories as a function of  $\lambda^*$, the ratio of particle radius to micropost radius.  Green (attraction to the post), red (equilibrium away from the post), and yellow (repulsion from the post throughout the field of view). The length of the bars indicate the probabilities of the observed particles trajectories within the $\lambda^*$ range.}
  \label{rep_fig_hist}
\end{figure}

\begin{figure}
\centering
  \includegraphics[scale=0.7]{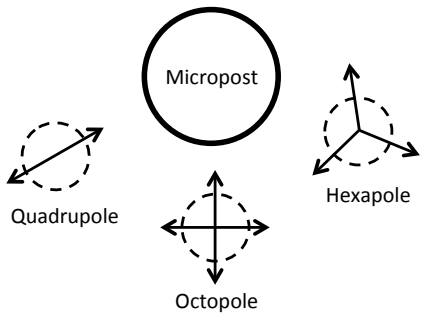}
  \caption[Preferred orientation of quadrupole, hexapole and octopole deformation around a micropost]{\textbf{Preferred orientation of quadrupole, hexapole and octopole deformation around a micropost.} Headed arrows indicate the axes of capillary rises on the particle. Quadrupole and octopus prefer their rise axes pointed toward the micropost, while hexapole prefers to point the valley between the two rise axes toward the micropost. }
  \label{rep_fig_modes}
\end{figure}

\begin{figure}
  \includegraphics[scale=0.9]{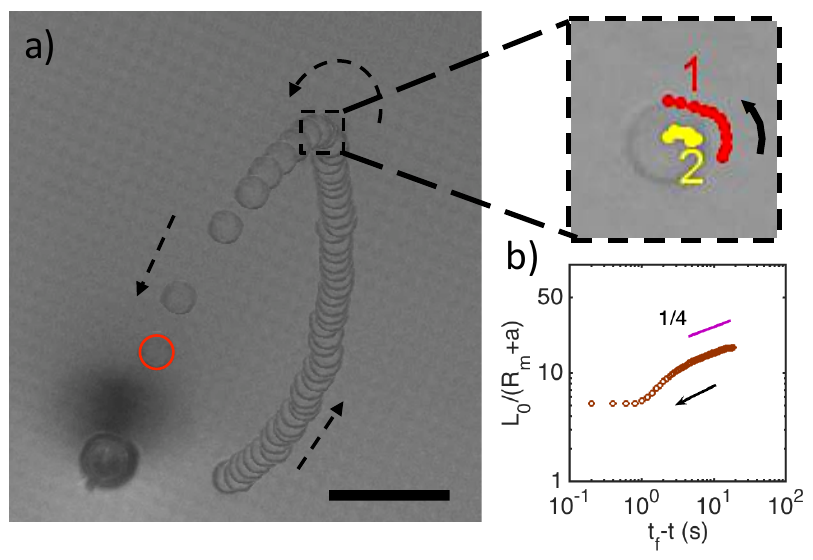}
  \caption{\textbf{Repulsion followed by rotation and attraction} (a) Time stamped image ($\Delta t$=2 s) of a microdisk ($2a$=25 $\mu$m) on a curved interface created by a 47 $\mu$m micropost. The microdisk is initially repelled by the curvature and migrates away from the micropost. The particle then rotates roughly $86^\circ$. Thereafter, it is attracted towards the micropost. Red circle indicates the final location of the disk. Inset: Particle rotation as marked by red dots (1) the mark on the particle, and yellow dots (2) indicate the center of the disk.  (b) $L_0/(R_m+a)$ vs $t_f-t$ for the latter part of trajectory, as disk approached the post.}
  \label{rep_fig_turnaround}
\end{figure}

We study microparticle migration on well-defined curved oil-water interfaces formed by pinning water around a micropost. We define $\lambda^*=\frac{a}{R_m}$; this parameter is the upper bound to $\lambda$ around any post.  Typical trajectories for disks of 25 $\mu$m in diameter on the curved interface molded by a 57 $\mu$m micropost are reported in Fig. \ref{rep_fig_baci}, for which $\lambda^*$=0.44. Three particles approach the micropost with distinct behaviors. The first disk migrated towards the micropost and stopped at center-to-center distance $L_0(t_f)$=50.1 $\mu$m from the post (Fig. \ref{rep_fig_baci}a). The second disk migrated at a much slower rate to the micropost and stopped at $L_0(t_f)$=128.0 $\mu$m (Fig. \ref{rep_fig_baci}b). The third disk migrated along curvature gradients all the way to the edge of the micropost (Fig. \ref{rep_fig_baci}c). Interestingly, upon the attachment of the third disk (green cross mark), the nearby disk (red triangle) was repelled at distances of a few particle radii by the incoming particle. It moved away from the incoming  particle by orbiting around the post, i.e. keeping a fixed distance from the post. \\
\indent There is evidence of competing modes when the trajectories are inspected for power law dependencies. Trajectories in Fig. \ref{rep_fig_baci} are shown in a log-log as insets. In these insets, the abscissa is $t_{f}-t$, and vertical axis is $L_0/(R_m+a)$, a normalized center-to-center distance which is unity for particles attached to posts. Far from the post, all three disks obey a power law of $\frac{1}{4}$, indicating a dominant quadrupolar mode. The trajectories deviate strongly from this power law for $L_0/(R_m+a) \sim 3$, or $L_0<$ 120 $\mu$m and $\lambda > 0.1$, consistent with contributions from other modes. These deviations differ from each other; two of the trajectories steepen, while the other plateaus at distances too far from contact to attribute to hydrodynamic interactions.  \\
\indent Experiments with larger disks (2a=150 $\mu$m) also reveal interesting behaviors. Images of the disks on planar interfaces are shown in Fig. ~\ref{rep_fig_largep}a and b. The disks are quite rough, with strong distortion apparent near their edges in the interferogram; the interface shape around the disk is randomly puckered near the particle's edge and weakly quadrupolar in the far field; an additional interferogram with random puckering near the disk and only higher order modes evident $\sim$ 2.5 radii from the disk in shown in Supplemental Material. We also show two example trajectories in Supplemental Material of such disks. A disk on an interface around a 600 $\mu$m post ($\lambda^*$ = 0.25) exhibits curvature attraction until contact, where on an interface around a 250 $\mu$m diameter micropost ($\lambda^*$ = 0.6), such a disk moved radially towards the post and stopped at some equilibrium location $L_0(t_f)$=314.7 $\mu$m, similar to the results above with the smaller disks. However, when such disks are placed on an interface around a micropost with diameter of 125 $\mu$m ($\lambda^*=$1.2), the disks are strongly repelled from the high curvature zones, migrating away from the micropost (Fig. \ref{rep_fig_largep}a) at rapid rates that diminish with distance from the post. \\
\indent The energy dissipated along these repulsive trajectories is shown versus $L_0$ in Fig. \ref{rep_fig_largep}c; these data, graphed against deviatoric curvature, suggest a linear relationship only far from the post (Supplemental Material); an inspection of $L_0$ versus $t$ suggests that these particles act as repulsive quadrupoles in the far field, i.e. that the particles have quadrupolar modes that are misaligned, and hence repelled from the high curvature regions, with $\alpha_2=\pi/2$.  The arrows in these figures indicate that the direction of migration is away from the post. \\
\indent We have conducted similar experiments with 25 $\mu$m disks around microposts of various diameters where $\lambda^*$ is 0.20, 0.44 and 0.53. We summarize our experimental findings for all of the disks in a histogram in Fig.~\ref{rep_fig_hist} and categorized these behaviors as attraction, equilibria or repulsion as a function of $\lambda^*$. The histogram shows systematic changes with $\lambda^*$. For $\lambda^*< 0.075$ all particles experience capillary curvature attraction and move without interruption along radial paths to the micropost. For moderate  $\lambda^*$, particles are either attracted all the way to the micropost or find equilibrium locations at distances of several  particle radii from the micropost. For larger $\lambda^*> 0.5$, particles either find equilibrium locations or are repelled from high curvature regions.  Finally, for the largest value of  $\lambda^*$ explored, all particles were repelled from the high curvature zones. \\
\indent We interpret these results within the context of (\ref{manymodenergy}). The contact line on each disk is determined by randomly distributed rough sites that pin the contact line. Thus, the amplitude of each mode can vary strongly from particle to particle in the same batch. Equation (\ref{manymodenergy}) indicates that each mode has a preferred orientation to minimize the energy on the interface. For example, for contact lines described by a quadrupolar mode, the energy is minimized when the quadrupolar rise axis aligns with the rise axis of the interface; this has indeed been reported for small cylindrical particles on curved fluid interfaces that excite strong quadrupolar distortions \cite{Cavallaro27122011}. For contact lines described solely by hexapolar modes or octopolar modes, particles would orient as shown in Fig. \ref{rep_fig_modes}. Particles experience a capillary torque to enforce this alignment; if all $\alpha_n$ could align so that their curvature capillary energy were negative,  the particles would migrate toward the post to minimize the capillary energy.  This migration would occur at different rates  owing to the different amplitudes of the modes $h_i$ and their differing dependencies on $L_0$. Such trajectories would give the steepest reduction in curvature capillary energy.\\
\indent However, random pinning implies not only random amplitudes, but also random phase angles of the various modes, so modes cannot co-align. Furthermore, when several modes are present, different modes dominate in different zones of the interface. Far from the micropost, the quadrupolar mode dominates the particle's rotational alignment and migration.  As the particle migrates toward the post, $\lambda$ increases, and higher order modes grow in importance. There are two scenarios in zones where these higher order modes compete. The particle may rotate, adjusting its angle in local equilibrium as it approaches the post. Or, there may be significant energy barriers to rotation owing to the rugged contact line, so particles cannot find an angle that allows particles to migrate toward the post. This latter case is particularly interesting, as it suggest that rough particles on curved interfaces may not rotate freely, but may have pinned orientations. For such pinned angles, the signs on the competing modes will likely differ.  Orientations can occur where attractive and repulsive terms can balance, defining states of mechanical equilibrium. However, as $\lambda$ approaches unity, all terms become important. For particles with trapped ``bad" alignments with contributions from many modes, energy can decrease only by moving \textit{away} from the post, i.e. they will be repelled from high curvature sites. \\
\indent We have observed a complex trajectory that further supports this interpretation. A 25 $\mu$m disk attached to an interface around a micropost of diameter 47 $\mu$m ($\lambda^*$ = 0.53). The particle was initially repelled  from the high curvature region, and migrated away from the post (Fig. \ref{rep_fig_turnaround}). However, at $L_0$= 309.5 $\mu m$ ($\lambda$ = 0.039) the particle suddenly rotated roughly $86 ^{\circ}$, and subsequently moved towards the post, obeying a 1/4 power law over the attractive segment of its motion. The disk then increased its speed as it neared the post, but stopped abruptly several particle radii from the post. This evidence suggests that the initial repulsion can be attributed to the trapped poorly aligned quadrupole. The rotation occurred at $\lambda$ where the higher order modes were negligible; the rotation angle of $86 ^{\circ}$ rotated the misaligned quadrupole into favorable alignment for curvature attraction. Finally, the power law in the subsequent attraction revealed the attractive quadrupolar curvature interaction (Fig. \ref{rep_fig_turnaround}b).  \\
\indent We can indeed construct simple deformation fields and energy landscapes that capture the main features of these experiments assuming a particle with two pinned modes, a hexapole and quadrupole, with the rise axis of quadrupole aligned with that of the hexapole.  If the particle attaches far from the post, the hexapolar mode can be neglected. The particle will rotate, aligning the quadrupolar rise axis with the rise axis of the interface and move to sites of high curvature.  As the particle approaches the post, the mis-aligned hexapole introduces an increasingly strong repulsive force. If rotation is impeded, the disk will find an equilibrium location which stops particle migration. If instead, the disk adsorbs to the interface near the micropost, both modes contribute and there will be a strong rotational barrier. Particles with misaligned modes can reduce their energy by moving away from the post (See Supplemental Material). We are exploring these scenarios using particle engineered to present such distortions in ongoing work.
%\begin{figure}
%\centering
%  \includegraphics[scale=1]{rep_energy_distance}
%  \caption{\textbf{Energy dissipation of particle trajectories from Fig.~\ref{rep_fig_edge_eqm_rep} and Fig.~\ref{rep_fig_largep}.} Energy dissipation versus distance from the center of micropost for (a) particle trajectories red, blue, and green colors correspond to Fig.~\ref{rep_fig_edge_eqm_rep} (a), (b) and (c) respectively and (b) particle trajectories from Fig.~\ref{rep_fig_largep} (c), which are repelled from high curvature region.}
%  \label{rep_fig_fig4}
%\end{figure}
\section{Conclusions}
 We have been developing the relationships between capillary energy landscapes around isolated particles and interface shape to develop new ways to direct particle migration and assembly. 
Contact line pinning plays an important role in determining such interactions; undulated three-phase contact lines create quadrupolar distortions to leading order. These distortions couple with the deviatoric curvature field of curved fluid interfaces.  The resulting curvature capillary energy drives small particles to sites of strong deviatoric curvature. In other studies, we have explored this phenomenon for a variety of microparticle systems, and have extended these arguments to understand microparticle migration on tense lipid bilayer vesicles, on which particles trace Brownian trajectories guided by curvature gradients owing to the weak tension in these systems \cite{Ningwei}.  Thus, the concepts we develop here can be adapted to apply to different physical systems.\\
\indent Here, we report that higher order modes in the interface shape and the particle sourced distortion provide greater complexity, with particles finding equilibrium locations far from the sites of highest curvature, or being repelled from high curvature locations. We have demonstrated these concepts using particles with random roughness to pin the contact lines. Pinned contact lines also occur on relatively smooth but chemically heterogeneous particles. Since such particles are ubiquitous, these phenomena are also likely of broad importance. 
\section{Outlook}
\begin{figure}
\centering
  \includegraphics[scale=0.7]{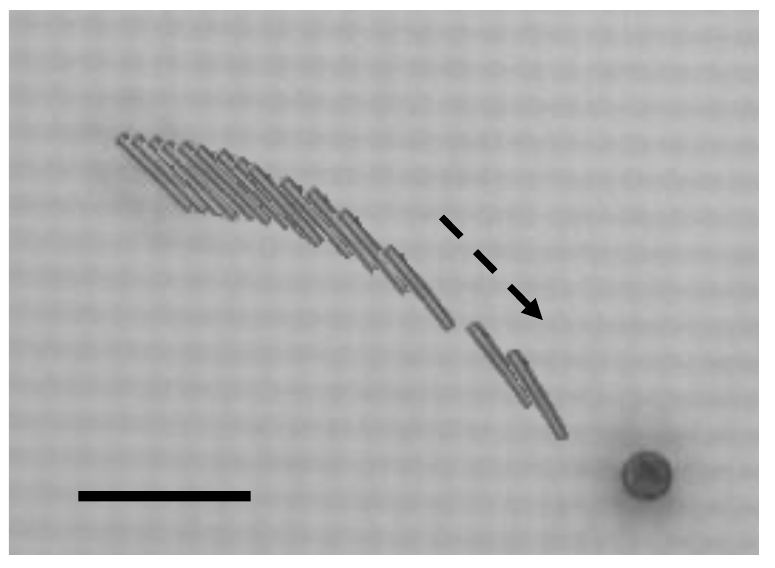}
  \caption{\textbf{Particle shape: A long microcylinder finds an equilibrium location with complex orientation}. Time stamped image ($\Delta t= 1.0 s$) of a microcylinder migration on a curved oil-water interface formed around a 28 $\mu m$ micropost. Scale bar is 100 $\mu$m.}
  \label{rep_fig_lu}
\end{figure}

Interface curvature has already been used to guide structure formation. For example, Ershov \textit{et al.} has exploited the coupling of the particle's quadrupolar distortion with the principal axes of curvature to form colloidal crystalline domains with quadrupolar symmetry from colloidal spheres \cite{Ershov}. Cavallaro \textit{et al.} has exploited this coupling and capillary curvature attraction to build complex structures around a square micropost \cite{Cavallaro27122011}. The building of such structures is a focus of ongoing work in our research group. Particles with differing shapes give important degrees of freedom; cylindrical particles show competition between pair quadrupolar interactions and curvature interactions in their assembly \cite{Lewand2009SM}. \\
\indent Curvature repulsion, discussed here, opens exciting new possibilities. To explore these phenomena, we have expanded the host interface locally to higher order modes in a multipole expansion. The particular expansion that we find is valid for our particular interface shape. However, such expansions can always be performed for interfaces with small slopes. Thus, this work suggests that interface shapes can be molded and tailored to emphasize particular modes in particular spatial regions. These interactions also have rich coupling with particle shape. For example, consider a high aspect ratio microcylinder on a curved interface formed around a very small micropost in Fig.~\ref{rep_fig_lu}. For reference, prior studies with small cylindrical microparticles around large posts report particles aligned with their major axes along principal axes. Here, the long microcylinder migrates radially along curvature gradients with its major  axis pointed toward the micropost, and stops at an equilibrium location far from the post with an orientation that positions its sharp edge toward the post, indicating the importance of details in the interface shape around the particle. This, too, is a focus of ongoing work. \\
\indent This research opens interesting new directions. In materials science, we might design curvature capillary energy wells to mold objects based on their symmetries far from the curvature source, thereby preventing trapped states that can occur for objects in contact with the posts.  In microrobotics, curvature attraction and curvature repulsion could provide new means of propelling microrobots along interfaces \cite{Wong2016}. It would be interesting to investigate limiting length scales; can we mold structure formation at the submicron scale?  Finally, we close by noting that this study falls within a rich and interesting class of problems in which a colloid is placed in soft matter, distorts the soft matter host, and creates an energy landscape around it. By molding the host-in this communication, the fluid interface, rich phenomena emerge. We and others have explored related phenomena for colloids and macromolecules confined in other hosts including colloids in nematic liquid crystals, colloids adhered to lipid bilayer vesicles, and proteins  on curved lipid bilayers \cite{Ningwei, IrisPNAS, Poulin21031997, doi:10.1021/la901719t, Lapointe20112009, Musevi?18082006}

\begin{acknowledgments}
We gratefully acknowledge Prof. Robert Carpick's group for the use of interferometer. This work was supported by the National Science Foundation (NSF) DMR-1607878.
%Materials Science and Engineering Center (MRSEC) Grant to University of Pennsylvania, NSF11-20901. 
%I.B.L. is supported by a Department of Education GAANN Grant P200A120246.
\end{acknowledgments}

%\bibliography{references}

\end{document}